# NONLINEAR METAMATERIALS FOR HOLOGRAPHY

Euclides Almeida[1], Ora Bitton[2] and Yehiam Prior[1]
[1]*Department of Chemical Physics, Weizmann Institute of Science, Rehovot 76100, Israel*
[2]*Department of Chemical Research Support, Weizmann Institute of Science, Rehovot 76100, Israel*
yehiam.prior@weizmann.ac.il
Euclides.almeida@weizmann.ac.il

**ABSTRACT**

A hologram is an optical element storing phase and possibly amplitude information enabling the reconstruction of a three dimensional image of an object by illumination and scattering of a coherent beam of light, and the image is generated at the same wavelength as the input laser beam. In recent years it was shown that information can be stored in nanometric antennas giving rise to ultrathin components. Here we demonstrate nonlinear multi-layer metamaterial holograms where by the nonlinear process of Third Harmonic Generation, a background free image is formed at a new frequency which is the third harmonic of the illuminating beam. Using e-beam lithography of multilayer plasmonic nanoantennas, we fabricate polarization-sensitive nonlinear elements such as blazed gratings, lenses and other computer-generated holograms. These holograms are analyzed and prospects for future device applications are discussed.

## INTRODUCTION

Holography was invented many decades ago[1,2] and was rapidly recognized as an effective means for storing and reconstructing images. By interference of a beam coming from the object and a reference coherent (laser) beam, phase and possibly also amplitude information has been stored in some media, typically a film which responds to the intensity of the light falling on it. The holographic method was developed in order to correct spherical aberration of electron lenses, but following the invention of the laser, optical holography [2,3] overshadowed the field of holography, eventually leading to applications such as volumetric data storage[4], optical tweezers [5] and 3D displays[6]. Other kinds of waves were also used in non-optical holography such as electron gas quantum holography[7] and plasmon holography[8].

In recent years, new storage elements have been proposed[9-11] using optical metamaterials which consist of (metallic) nanoantennas that typically encode phase information for reconstructing images (phase holograms). Metamaterials are periodically nanostructured artificial materials which can manipulate light-matter interactions on spatial dimensions smaller than or comparable to the wavelength of light[12]. The optical response of these metamaterials can be engineered to exhibit unusual optical phenomena such as negative refraction[13,14] or optical cloaking[15], and can lead to novel applications. Metasurfaces are ultrathin, quasi-2D metamaterials made of metallic or dielectric nanostructures which can locally control the phase, amplitude or the polarization state of light waves propagating through or reflected from them[16-21]. Such control is the working principle behind their application as holographic metasurfaces[22]. Three dimensional metamaterials[23,24] were also used[9]. So far, the vast majority of metasurfaces have been designed to operate in the linear regime. Therefore, while they shape the wavefront of light, they cannot alter its frequency. More recently, the concept of phased metasurfaces was extended to the nonlinear (NL) regime[25-28], enabling both coherent generation and manipulation such as beam steering, and lensing of light beams. Nonlinear phase control has been demonstrated for Second Harmonic Generation (SHG) in arrays of metallic split-ring resonators[25], Third Harmonic Generation (THG) in cross-shaped metallic nanoparticles for circularly polarized light[26] and four-wave mixing in metallic thin films[28,29].

Plasmonic metasurfaces [9-11] have been used for computer generated holograms[30], where a target image is digitally processed and the phase pattern of the hologram is calculated using numerical methods for light propagation/diffraction. The image is reconstructed by a reading laser beam that illuminates the storage medium. In standard computer-generated holography, the image is formed at the same wavelength of the reading laser beam.

Here we build on the concept of nonlinear phase control in plasmonic metasurfaces, and demonstrate THG holograms where the image is formed at a wavelength different from the reading beam, as illustrated in Fig. 1. We demonstrate computer-generated holograms stored in multilayered plasmonic metasurfaces, and exhibit polarization as an additional control parameter. The high density storage enabled by sub-micron nanoantennas, especially in multilayered structures, may lead to holograms with very high resolution.

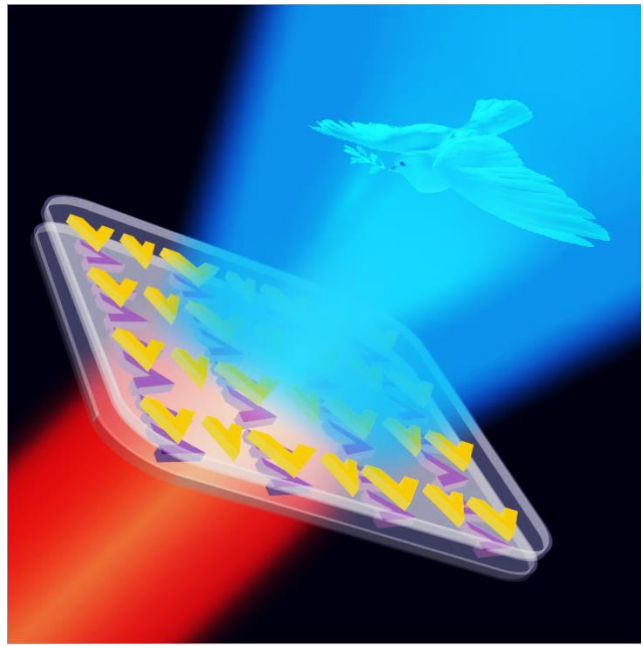

Figure 1: Artist's concept of a multilayer NL metamaterial hologram. The phase information of an object is encoded in computer generated nanoantennas in the multi-layer metamaterial. After illumination with a suitably polarized infrared laser, the image of the object is reconstructed at a frequency that is the third-harmonic of the incoming beam.

## RESULTS

**Antenna design**

As our basic nanostructured element we chose linearly polarizable V-shaped gold antennas as depicted in Fig. 2a. Metallic nanostructures are attractive candidates for efficient harmonic generation due to the high nonlinearities of metals[31], and the near-field enhancement at plasmonic resonances which can boost frequency conversion by orders of magnitude[25]. For symmetric V-shaped antennas, the two parameters that we change are the length of the arms and the angle between them. As these parameters are changed, the plasmonic resonances are tuned across the near-infrared spectrum. Near resonance, the electronic cloud of the nanoantennas is driven by a phase-shifted effective field $a(\omega)E_1 e^{i\varphi(\omega)}$, where $E_1$ is the incoming light field, $a(\omega)$ is the near-field enhancement and $\varphi(\omega)$ is the shape-dependent phase-shift imposed on the incoming beam. Approximating the NL nanoantennas as point dipoles of effective third-order NL susceptibility $\chi_{eff}^{(3)}$, the third-order polarization (oscillating at 3ω) induced in the dipole is given by

$$P^{(3)}(3\omega) \propto \int \chi_{eff}^{(3)}(3\omega,\omega)\left[a(\omega)E_1 e^{i\varphi(\omega)}\right]^3 \tag{1}$$

Therefore, if $\varphi(\omega)$ is the phase-shift of the fundamental beam, the phase shift to the third-harmonic field $E_3$ will be $3\varphi(\omega)$ where an additional relative phase shift may be added[28] depending on the resonant nature of $\chi_{eff}^{(3)}$. Fig. 2 provides the information about the antennas and their phase control. Fig. 2a depicts the antenna and the dimensions of the relevant parameters (arm length and angle); Fig. 2b depicts Finite Differences Time Domain (FDTD) calculations of the linear plasmonic resonance transmission intensity for various combinations of arm length and angles. It is clearly seen that the resonance peak moves to longer wavelengths for longer arm length and smaller angles.

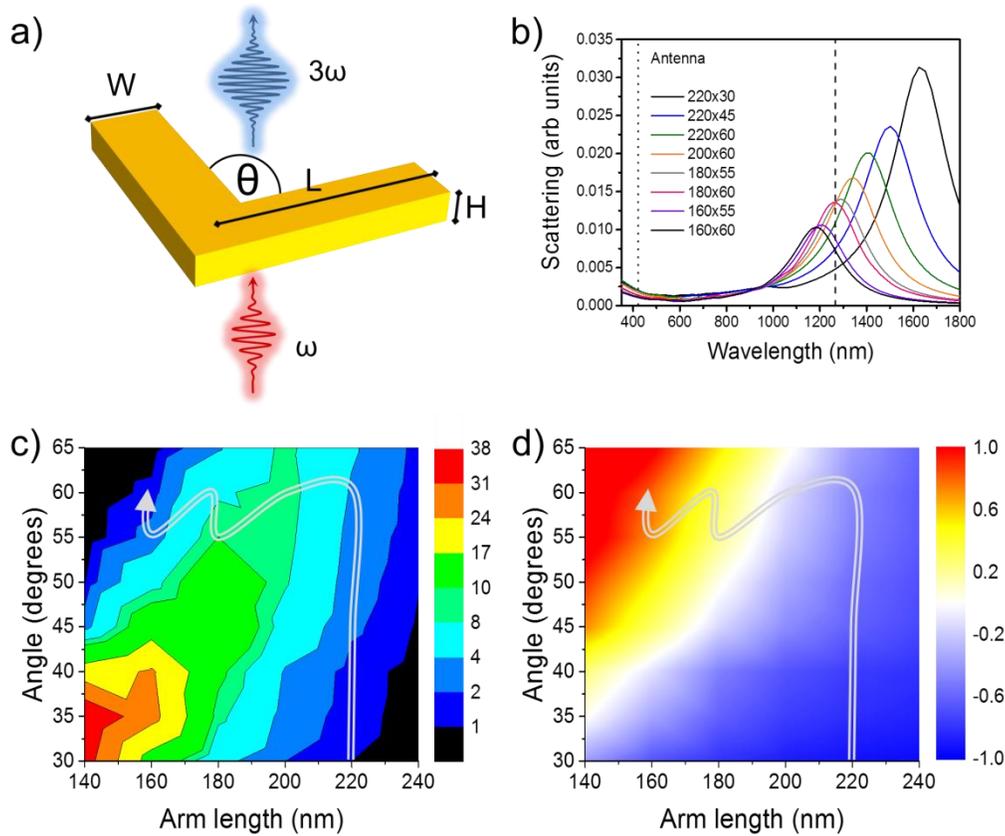

Figure 2 – Design of phase-controlled nonlinear nanoantennas. a) Dimensions of the V-shaped gold antennas for local control of the phase of the nonlinearity. The free parameters are the arm length L and the angle $\theta$ between the arms. H= 30 nm and W=40 nm are constant for all antennas. b) Representative linear scattering spectra of some of the antennas (Lx $\theta$), the input light was polarized along the bisecting axis of the V-shaped antennas. The dashed and dotted lines indicate the free-space wavelengths of the fundamental ($\lambda_F = 1266$ nm) and third-harmonic ($\lambda_F = 422$ nm) beams respectively. Amplitude (c) and phase (d) map of the third harmonic signal. The color code in the phase map is represented in units of $\pi$ and the nanoantennas used in the devices lie along the path drawn in the maps.

However, to calculate the phase holograms we need phase information at the THG wavelength. The calculated amplitude and phase, Non Linear FDTD (NL-FDTD), of the THG generated at $3\omega = 422 nm$ are presented in Figs. 2c and 2d. We used a realistic parameter range that is enough to generate a $2\pi$ relative phase shift. The wavy arrows in both panels depict a path for which the phase varies by up to the required $2\pi$ with a minimal variation of intensity, certainly much less than an order of magnitude. These linear antennas are capable of sweeping the $2\pi$ phase range when both the fundamental and the third harmonic beams are of the same polarization. Furthermore, the antennas emit directional dipole radiation which provides the basis for our polarization controlled holography, namely, the intensity of the THG of the V-shaped antennas is strongly polarization-dependent, and the signal is almost negligible for a horizontally polarized fundamental beam.

The basic set-up of a two-layer polarization sensitive nonlinear NL blazed grating is shown and discussed in Supplementary Figure 1. Most importantly – since these nanoantennas do not show any plasmonic resonances at the third-harmonic frequency, the phase is not changing upon propagation through the sample, thus enabling us to introduce multilayer structured phase holograms.

**Sample fabrication**

The samples are prepared by multilayer e-beam lithography on a borosilicate glass substrate (details are given in the Methods section). A 180 nm thick silica layer is deposited by plasma-enhanced chemical vapor deposition (PECVD) on the silica substrate. Then the desired design of 30 nm thick antennas is patterned into the silica layer by e-beam lithography, the process is then repeated for additional layers where the deposited silica layer serves as a dielectric spacer between two adjacent active nanoantenna layers.

**Computer Generated Metamaterial Holograms (CGMH)**

Several different types of metamaterial holograms were fabricated; all of them nonlinear where the wavelength of the illuminating input beam is 1266 nm and the image are formed at the third harmonic at 422nm. Different polarizations were utilized for different images, and each image was encoded into a single layer in a multilayer composite metamaterial hologram. In some cases (not shown here) more than one polarization coded image was stored in a layer. Some of the holograms were designed such that the each image was formed at a different distance, thus laying the foundation to 3D capabilities.

Fig. 3 depicts two different double layer structures with different holograms embedded in each layer for each structure. For each layer within the multi-layer metamaterial, a separate phase hologram was computer generated to yield the desired far field image for the proper polarization. For one structure the Hebrew letters Aleph and Shin are generated for vertical and horizontal polarizations respectively and for the second, images of happy and sad smiley faces are generated. Note that each image is recreated only by the properly polarized input beam, and emanates from a single layer of nanoantennas within the metamaterial. Fig. 3e depicts an enlarged section of the multilayer structure, demonstrating the stamping accuracy of the different layers

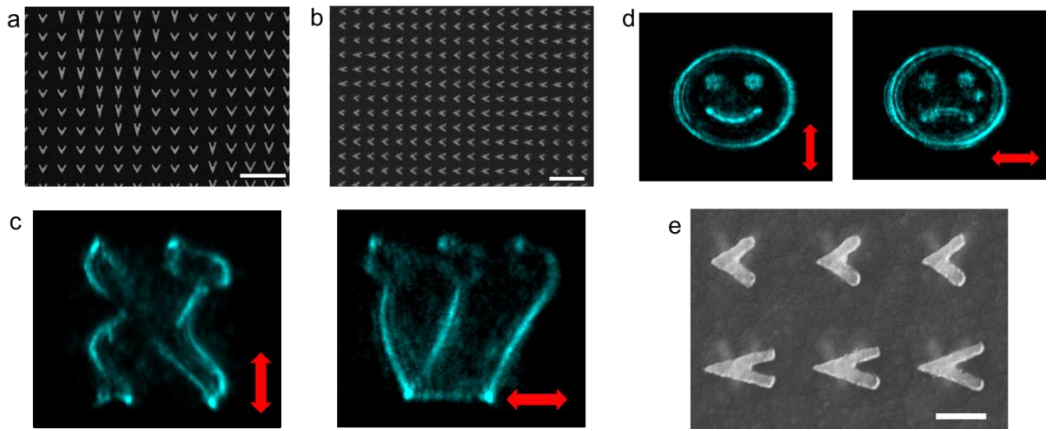

Figure 3 – Nonlinear metamaterial holograms. SEM images of the first (a) and the second layer (b) of a 3D nonlinear hologram which projects holographic images of the Hebrew letters Alef and Shin (c) for vertically or horizontally polarized input beams, respectively. (d) Nonlinear holograms of smiley and sad faces for vertically or polarized fundamental beams respectively. (e) SEM image of two typical overlaid layers of nanoantennas. The scale bars are 1 μm (a,b), 200 nm in (e)

These nonlinear phase-holograms work in transmission and at the THG frequency there is no background noise as is the case for linear transmission holograms where the image is at the same frequency as the illuminating beam.

As a further demonstration of the resolution of these metamaterial holograms we computer generated holograms depicting the famous line-drawings of the Dove of Peace and the Owl by Pablo Picasso. The holograms were generated for orthogonal polarizations at a focal plane that is 100 microns from the metasurface. Fig. 4 shows the images at the pre-designed target plane, where clear images are visible, and at the surface of the hologram where they are not.

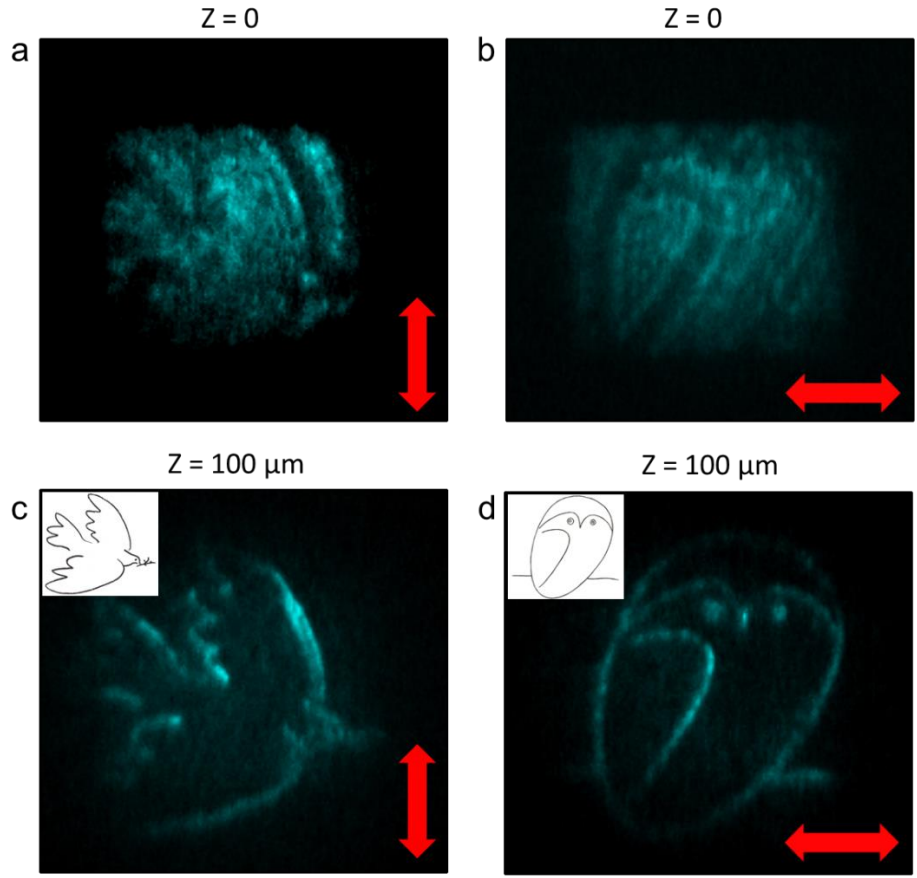

Figure 4: Holographic reconstruction of famous line drawings by Spanish artist Pablo Picasso (1881-1973). Blurred NL images at the surface of the hologram for vertical (a) and horizontal (b) polarizations. (c,d) Designed NL images at the correct distance of 100 μm from the surface. The width of the lines in the images is about 3 μm.

The next two examples of polarization dependent holograms depict images formed at different focal distances, three such distances in one case, and two in another. Following up on the choice of the Picasso drawings, we computer generated a hologram indicating the name of the Metropolitan Museum in New York (MET), where each of the three letters was designed to be generated at different focal distance and for a different polarization (0º, 45º and 90º respectively). The result is depicted in Fig. 5. To further illustrate the sensitivity to the correct polarization, images generated with the 'wrong' polarization at the correct focal distance are also shown.

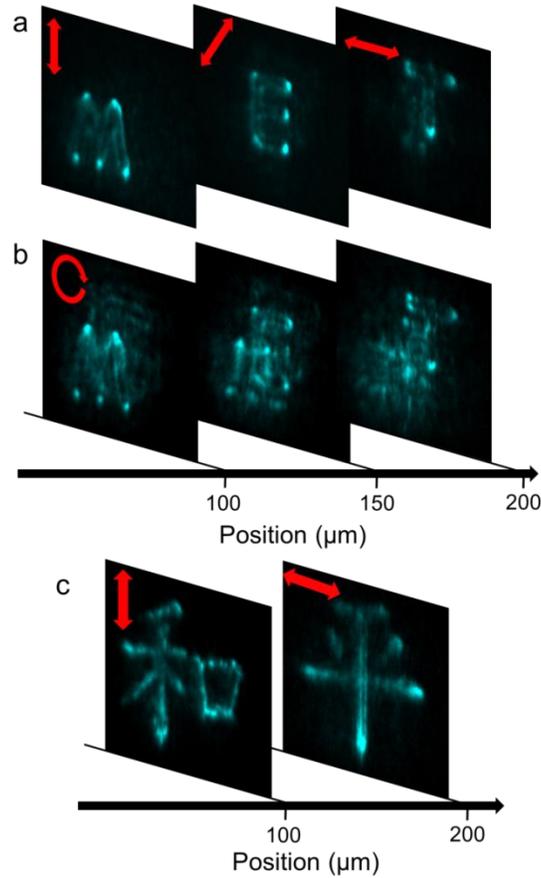

Figure 5: Nonlinear holograms displaying three-dimensional features. a) NL images of a three-layer hologram displaying the word MET at different planes for light polarized at 0, 45 and 90 degrees. b) NL images for input circularly polarized light. In a given plane, all the letters are visible but two of them appear unfocused. (c) Chinese characters for "Peace and harmony".

The last in this series of demonstrated metamaterial holograms are the Chinese characters for 'Peace and Harmony' (Fig. 5c) where greater detail and better accuracy are required. The images are recreated by the proper polarizations at the different focal distances.

In Supplementary figure 2 we demonstrate how different polarizations give rise to three different but related images, and in Supplementary movie 1 the continuous scanning of the input polarization creates a dynamic hologram that gives the impression of moving wings of a fan.

The direct implementation of the ability to steer the beam in a polarization dependent manner is manifested in the design of nonlinear metalenses. A nonlinear Fresnel-like multilayer metalens focuses the TH radiation into two different focal points depending on the input polarization of the fundamental beam (Fig. 6). The phase profile necessary to obtain light focusing is given by $\phi(r) = 2\pi/\lambda_{TH}\sqrt{r^2 + f^2}$, where $\lambda_{TH}$ is the free-space wavelength of the third-harmonic beam and $f$ is the focal distance. The metamaterial focuses the TH at $f = $ 1mm for vertical polarization and at $f = $ 500 µm for horizontal polarization into nearly diffraction-limited spots. The lenses are intrinsically nonlinear and work only for the signal that is coherently generated in situ. An incoming beam of the frequency identical to that of the THG beam will not be focused, nor is the incoming fundamental beam focused by this unique lens. The actual beam propagation through the focal regions of such metalenses with focal distances of 1 and 0.5 mm are shown in Supplementary Movies 2, 3.

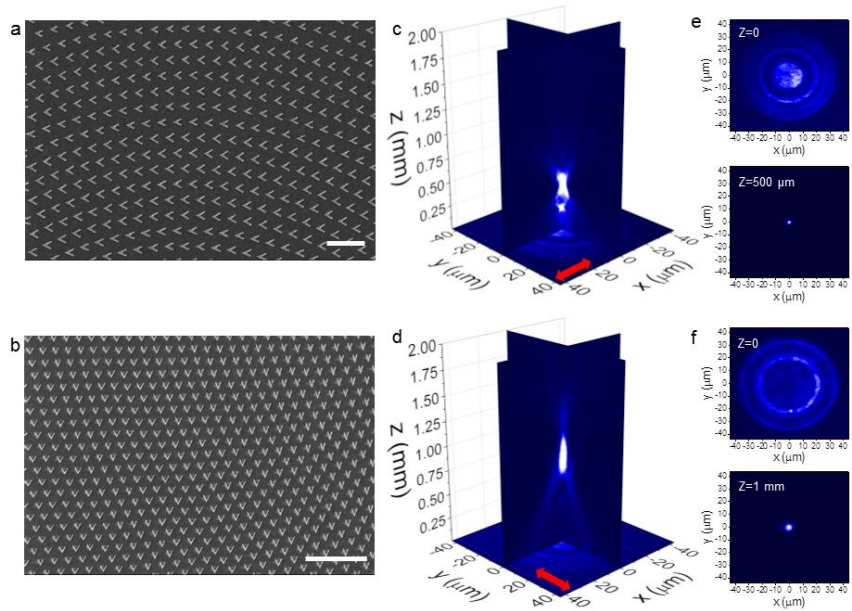

Figure 6 – A polarization controlled nonlinear lens. SEM images of the first (a) and the second (b) layers of a Fresnel lens made by a 3D nonlinear metamaterial. The scale bars are 1 and 2 µm in a) and b) respectively. Microscopic images of the third harmonic signal away for the surface of the metamaterial device for a horizontally (c) or vertically (d) polarized fundamental beam. The first active layer (a) focuses the TH radiation around the focal point $z = $ 500 µm (c), while the second layer focus the light at $z = $ 1 mm (d). Microscopic images at $z = $ 0 and $z = $ 500 µm for horizontal polarization (e) and $z = $ 0 and at $z = $ 1 mm for vertical polarization (f). The Gaussian RMS widths at $z = $ 500 µm and $z = $ 1 mm are $\sigma = $ 1.2 and 2.2 µm respectively, close to the theoretical diffraction limited values $\sigma = $ 1.1 and 2.2 µm.

**DISCUSSION**

The ability to design and fabricate nonlinear metamaterials with local control of the nonlinear phase in all spatial dimensions enable the realization of efficient functional NL devices with additional degrees of freedom and may have applications in volumetric data storage. In this work, we demonstrated polarization sensitive holograms that were designed for generating the third harmonic of a specific input frequency. Ours are phase-only holograms, but in principle, the amplitude may also be controlled at the individual element level. While metasurfaces can also provide polarization sensitivity, the volumetric metamaterials enable higher storage density, which is crucial for realization of efficient nonlinear devices. Moreover, since the individual element can be much smaller than the optical wavelength, the maximum data density can be much higher than with standard films where the resolution is limited by optical diffraction.

The holographic method demonstrated in this work differs from the conventional optical holograms in the sense that in the present case the image is formed at a different frequency. This enables the realization of transmission holograms with low background. By changing the input polarization, different images are revealed, either at the same plane or at different planes, thus enabling the formation of dynamic three dimensional images which are 'moving' in response to changing polarization.

In summary, we demonstrate a novel method of volumetric storage of optical information using nonlinear 3D metamaterial devices. By encoding the NL phase in three-dimensions, we build computer generated polarization sensitive nonlinear holograms as well as optical elements such as lenses and blazed gratings. In this work we converted infrared incoming beam to visible-range image, but the concept remains the same for other spectral ranges as well. These high-density NL Holographic 3D metamaterials may find applications as data storage devices and integrated nonlinear photonics.

# METHODS

## Numerical simulations

The linear and nonlinear optical responses were calculated by the FDTD method using the commercial software Lumerical FDTD Solutions[32]. In all the simulations, the dimensions of the mesh around the nanoantennas were set to dx=dy=dz=5 nm, and perfectly matched layers were added in all dimensions to avoid reflections. The optical constants of Au and Cr were extracted from the CRC tables, while the optical constants of $SiO_2$ were taken from Palik[33]. The linear scattering of the nanoantennas was calculated using a total-field/scattered field source. The third-order nonlinear susceptibility of gold was set to $\chi^{(3)}=10^{-19}$ $m^2/V^2$. The nonlinear light source was a linearly polarized transform-limited 60-fs long pulse centered at 1266 nm with amplitude $5\times10^8$ V/m. The complex electric field of the third-harmonic signal emitted in the forward direction is recorded on a y-normal plane, from which we extract the relative phase of the third harmonic signal. The amplitude of the THG is calculated by integrating the z component of Poynting vector on a power detector placed in the far-field.

## Sample Fabrication

The multilayer samples were fabricated using e-beam lithography. A 150 μm thick microscope cover slip was used as the substrate. To avoid electron charging on the substrate, a 3 nm thick chromium layer was initially deposited on the substrate by e-beam evaporation. A 180 nm silica layer was deposited by PECVD on top of the Cr film. The patterns were exposed by e-beam lithography (RAITH e_Line Plus) on a 125 nm thick resist (950k PMMA A2). We used accelerating voltage of 30 kV and a beam current of 30pA. Alignment marks were exposed at every e-beam exposure step and these were used for aligning the pattern of the next (top) layer. By using alignment marks we achieved overlay accuracy between each two adjacent layers as high as 10nm. After the sample was developed in MIBK:IPA (1:3), the silica layer was etched by Inductively Coupled Plasma (ICP SPTS). We used $CF_4$ gas, 50 sccm flow, 400W ICP power and 50W platten power. Etch time of 20 s results in about 30 nm etch depth in the SiO2 layer. A 2 nm thick chromium adhesion layer and a 30 nm thick gold layer were then e-beam evaporated on the etched patterns. The resist is then lifted off in acetone leaving metallic patterns stuck in the SiO2 layer. A 180 nm thick silica layer is deposited again on the metallic pattern by PECVD. The whole fabrication process is repeated for the additional layers. The total lateral size of the structured sample is 80 microns, and to minimize near field coupling, the center-to-center distance between the nanoantennas was maintained to be 410 nm. Further information is provided in Supplementary Method 1

**Nonlinear measurements**

An Optical Parametric amplifier (OPA) pumped by an amplified Ti:Sapphire laser was used as the light source. The OPA delivered 60 fs long pulses, centered at 1266 nm and with repetition rate of 1 kHz. An achromatic half-waveplate (Casix, WPZ1315) was used to rotate the polarization of the beam. The fundamental beam was weakly focused on the sample by a 1 m lens to a waist of approximately 100μm. The average power was set to 380μW. The TH signal was collected by an objective lens of numerical aperture 0.44 and filtered by bandpass and shortpass filters (Semrock, brightline 405/150, Semrock EdgeBasic SWP785, Thorlabs FESH0750). The images of the holograms and the lenses were projected onto an EMCCD camera (Andor, iXon$^{EM}$+ 885) using another lens of focal distance 200 mm. A flip mirror was used to switch the signal between the EMCCD and a spectrometer (Jobin-Yvon, Triax 180 coupled to a CCD, Symphony, Jobin Yvon) for spectral analysis. The k-space projection of the NL blazed grating was done by imaging the back focal plane of the objective onto the CCD using a 50 mm lens.

**Hologram design**

The computer generated phase holograms were designed using the point source method. In this method, each pixel $(x'_i, y'_i)$ in the two-dimensional 'object' is a point source of spherical light waves and the phase element at the hologram is calculated as the linear superposition of the electric fields emitted by all the point sources of the image. The holographic image is projected on the CCD camera and digitally inverted to obtain the image as seen by an observer looking from the sample to the image.

# REFERENCES

1. Gabor, D. A new microscopic principle. *Nature* **161**, 777 (1948).
2. Leith, E. N. & Upatnieks, J. Reconstructed Wavefronts and Communication Theory*. *J. Opt. Soc. Am.* **52**, 1123-1130, doi:10.1364/JOSA.52.001123 (1962).
3. Denisyuk, Y. N. On the reflection of optical properties of an object in the wave field of light scattered by it. *Doklady Akademii Nauk SSSR* **144**, 1275 (1962).
4. Heanue, J. F., Bashaw, M. C. & Hesselink, L. Volume Holographic Storage and Retrieval of Digital Data. *Science* **265**, 749-752, doi:DOI 10.1126/science.265.5173.749 (1994).
5. Dufresne, E. R. & Grier, D. G. Optical tweezer arrays and optical substrates created with diffractive optics. *Rev Sci Instrum* **69**, 1974-1977, doi:Doi 10.1063/1.1148883 (1998).
6. Blanche, P. A. *et al.* Holographic three-dimensional telepresence using large-area photorefractive polymer. *Nature* **468**, 80-83, doi:10.1038/nature09521 (2010).
7. Moon, C. R., Mattos, L. S., Foster, B. K., Zeltzer, G. & Manoharan, H. C. Quantum holographic encoding in a two-dimensional electron gas. *Nat Nano* **4**, 167-172, doi:http://www.nature.com/nnano/journal/v4/n3/suppinfo/nnano.2008.415_S1.html (2009).
8. Dolev, I., Epstein, I. & Arie, A. Surface-Plasmon Holographic Beam Shaping. *Phys Rev Lett* **109**, doi:Artn 20390310.1103/Physrevlett.109.203903 (2012).
9. Larouche, S., Tsai, Y. J., Tyler, T., Jokerst, N. M. & Smith, D. R. Infrared metamaterial phase holograms. *Nat Mater* **11**, 450-454, doi:10.1038/NMAT3278 (2012).
10. Ni, X. J., Kildishev, A. V. & Shalaev, V. M. Metasurface holograms for visible light. *Nat Commun* **4**, doi:Unsp 280710.1038/Ncomms3807 (2013).
11. Huang, L. L. *et al.* Three-dimensional optical holography using a plasmonic metasurface. *Nat Commun* **4**, doi:Artn 280810.1038/Ncomms3808 (2013).
12. Cai, W. & Shalaev, V. *Optical Metamaterials, fundamentals and application*. (Springer-Verlag 2010).
13. Veselago, V. G. Electrodynamics of Substances with Simultaneously Negative Values of Sigma and Mu. *Sov Phys Uspekhi* **10**, 509-&, doi:Doi 10.1070/Pu1968v010n04abeh003699 (1968).
14. Pendry, J. B. Negative refraction makes a perfect lens. *Phys Rev Lett* **85**, 3966-3969, doi:DOI 10.1103/PhysRevLett.85.3966 (2000).
15. Cai, W. S., Chettiar, U. K., Kildishev, A. V. & Shalaev, V. M. Optical cloaking with metamaterials. *Nat Photonics* **1**, 224-227, doi:10.1038/nphoton.2007.28 (2007).
16. Bomzon, Z., Kleiner, V. & Hasman, E. Pancharatnam-Berry phase in space-variant polarization-state manipulations with subwavelength gratings. *Opt Lett* **26**, 1424-1426, doi:Doi 10.1364/Ol.26.001424 (2001).
17. Yu, N. F. *et al.* Light Propagation with Phase Discontinuities: Generalized Laws of Reflection and Refraction. *Science* **334**, 333-337, doi:DOI 10.1126/science.1210713 (2011).
18. Ni, X. J., Emani, N. K., Kildishev, A. V., Boltasseva, A. & Shalaev, V. M. Broadband Light Bending with Plasmonic Nanoantennas. *Science* **335**, 427-427, doi:DOI 10.1126/science.1214686 (2012).
19. Lin, D. M., Fan, P. Y., Hasman, E. & Brongersma, M. L. Dielectric gradient metasurface optical elements. *Science* **345**, 298-302, doi:DOI 10.1126/science.1253213 (2014).
20. Kildishev, A. V., Boltasseva, A. & Shalaev, V. M. Planar Photonics with Metasurfaces. *Science* **339** (2013).
21. Yu, N. F. & Capasso, F. Flat optics with designer metasurfaces. *Nat Mater* **13**, 139-150, doi:Doi 10.1038/Nmat3839 (2014).
22. Genevet, P. & Capasso, F. Holographic optical metasurfaces: a review of current progress. *Rep Prog Phys* **78**, doi:Artn 02440110.1088/0034-4885/78/2/024401 (2015).

23	Liu, N. *et al.* Three-dimensional photonic metamaterials at optical frequencies. *Nat Mater* **7**, 31-37 (2008).
24	Zhao, Y., Belkin, M. A. & Alu, A. Twisted optical metamaterials for planarized ultrathin broadband circular polarizers. *Nat Commun* **3** (2012).
25	Segal, N., Keren-Zur, S., Hendler, N. & Ellenbogen, T. Controlling light with metamaterial-based nonlinear photonic crystals. *Nat Photonics* **9**, 180-184, doi:Doi 10.1038/Nphoton.2015.17 (2015).
26	Li, G. X. *et al.* Continuous control of the nonlinearity phase for harmonic generations. *Nat Mater* **14**, 607-612, doi:Doi 10.1038/Nmat4267 (2015).
27	Wolf, O. *et al.* Phased-array sources based on nonlinear metamaterial nanocavities. *Nat Commun* **6**, doi:Artn 766710.1038/Ncomms8667 (2015).
28	Almeida, E., Shalem, G. & Prior, Y. Subwavelength Nonlinear Phase-Control and Anomalous Phase-Matching in Plasmonic Metasurfaces. *Nat Comm (In Press) and arXiv1505.05618* (2015).
29	Almeida, E. & Prior, Y. Rational design of metallic nanocavities for resonantly enhanced four-wave mixing. *Sci. Rep.* **5**, 10033 (2015).
30	Tricoles, G. Computer Generated Holograms - an Historical Review. *Appl Optics* **26**, 4351-4360 (1987).
31	Kauranen, M. & Zayats, A. V. Nonlinear plasmonics. *Nat Photonics* **6**, 737-748, doi:Doi 10.1038/Nphoton.2012.244 (2012).
32	Lumerical. Lumerical solutions, Inc. http://www.lumerical.com/tcad-products/fdtd. (2014).
33	Palik, E. D. *Handbook of Optical Constants of Solids* (Academic press, San Diego, California, 1998).

**ACKNOWLEDGEMENTS**

This work was funded, in part, by the Israel Science Foundation, by the ICORE program, by an FTA grant from the Israel National Nano Initiative, and by a grant from the Leona M. and Harry B. Helmsley Charitable Trust. Discussions with Guy Shalem, Roy Kaner, Yaara Bondy, Yael Blechman and Basudeb Sain are gratefully acknowledged. The help of Ishai Sher with the graphics is appreciated

**AUTHORS CONTRIBUTION STATEMENT**

EA and YP conceived the idea and interpreted the physical results. EA performed the experimental work and the numerical simulations. EA and OB fabricated the samples. All authors contributed to the writing of the paper.

**COMPETING FINANCIAL INTERESTS**

The authors declare no competing financial interests.

# Supplementary information for

# Nonlinear Metamaterials for Holography

Euclides Almeida[1], Ora Bitton[2] and Yehiam Prior[1]
[1]*Department of Chemical Physics, Weizmann Institute of Science, Rehovot 76100, Israel*
[2]*Department of Chemical Research Support, Weizmann Institute of Science, Rehovot 76100, Israel*
*Yehiam.prior@weizmann.ac.il*
*Euclides.Almeida@weizmann.ac.il*

**Supplementary Figure 1:** Two-layer polarization sensitive nonlinear blazed grating

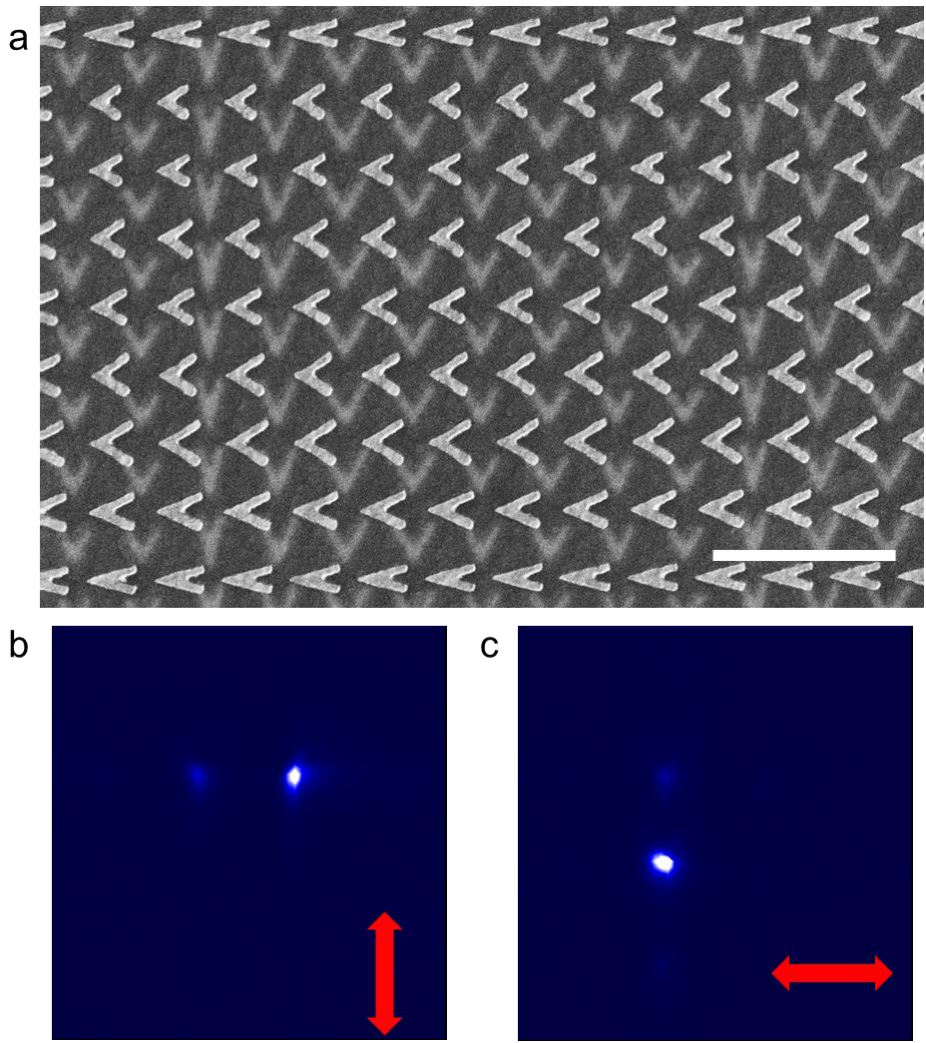

We designed a two-layer NL blazed grating which steers the TH beam into different planes. The SEM image is shown in (SF 1a). The scale bar is 1 µm. The top and bottom layers were shifted on purpose. In SF 1b-c, we show a k-space projection of the NL signal diffracted by the blazed gratings onto a CCD camera. An incoming y-polarized fundamental beam steers the third-harmonic along the horizontal direction (SF 1b), while an x-polarized incoming beam generates a TH signal which is steered along the vertical direction (SF 1c). Within a single unit cell, the $2\pi$ phase shift necessary to build an efficient blazed grating can be achieved by the nonlinear nanoantennas as shown in figure 1d in the main text. The zeroth order of diffraction as well as the negative and higher orders are much weaker than the $1^{st}$ order, which indicates that the phase shift from cell to cell is close to $2\pi$ as predicted by NL-FDTD calculations. The beam steering angle measured after calibration is $8.25 \pm 0.30$ degrees. As discussed in [1], the beam steering angle can be obtained analytically according to the anomalous phase matching condition, which for THG can be expressed as $k_{THG}^a = 3k_1 + d\Phi_{THG}/dx$, where $k_1$ is the wavevector of the fundamental beam and $d\Phi_{THG}/dx$ is the phase gradient of the third-harmonic beam along the interface. The anomalous phase matching condition is a manifestation of momentum exchange between the interface and all beams participating in the nonlinear process. For our blazed grating, the angle calculated according to the anomalous phase matching condition is 8.10 degrees, in good agreement with the experimental angle. The beam steering and light focusing capabilities of NL phase control may be a strategy to spatially filter the fundamental and nonlinear beams, and can be useful in nonlinear spectroscopy such as CARS for example.

**Supplementary Figure 2:** Dynamic nonlinear hologram

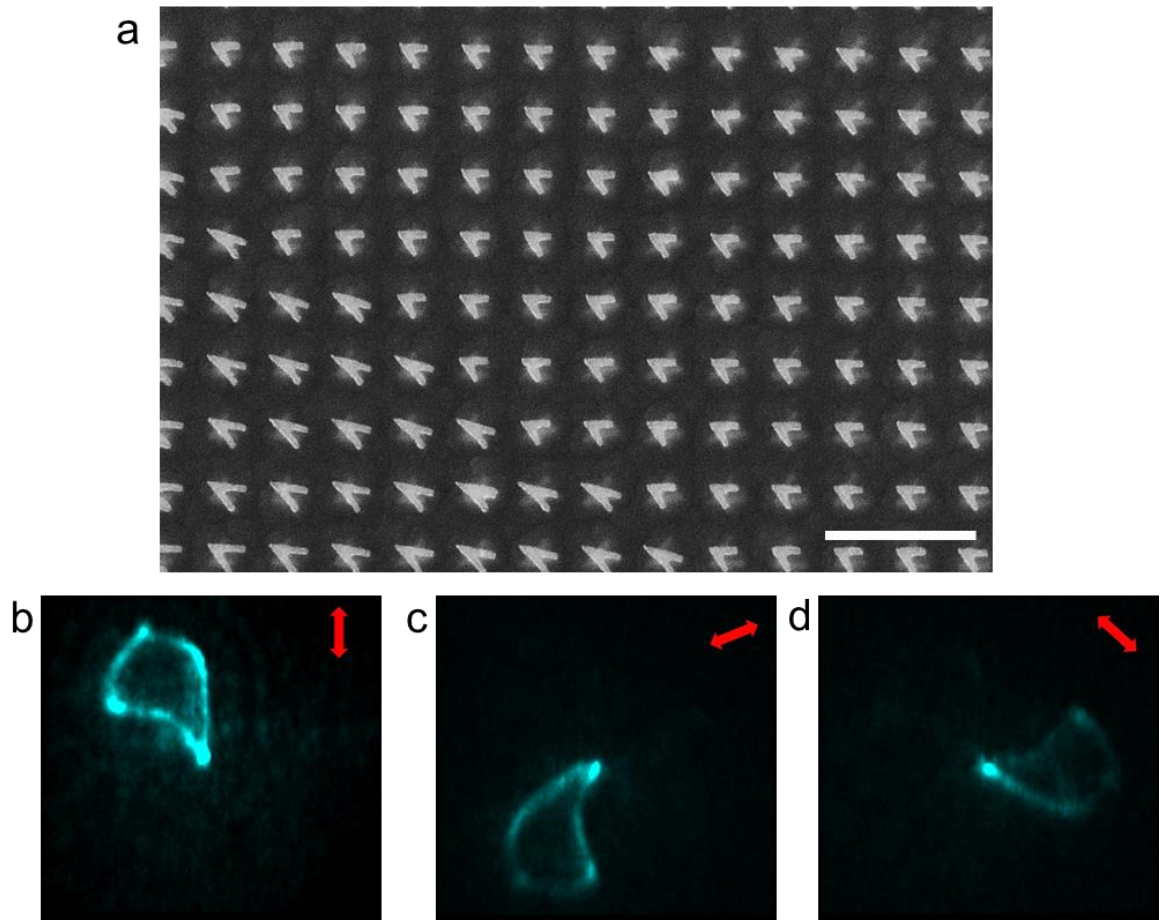

Supplementary Figure 2 depicts a dynamic nonlinear hologram build with a three layer metamaterial. The SEM image of a section of the third layer of the device is shown in Supplementary Figure 2a. The scale bar is 1 μm. In Supplementary Figures 1b-d, we show the holographic NL image for three different polarization states of the fundamental beam. For each polarization state shown, one of the layers of the metamaterial is active. The V-shaped gold nanoantennas are rotated by 60° from one layer to next. A fan-shaped movie can be played when the polarization is continuously rotated, as can be seen in Supplementary Movie 3.

**Supplementary Movie 1:** A dynamic nonlinear hologram

**Supplementary Movie 2:** polarization-controlled nonlinear metalens with focal distance of 1 mm

**Supplementary Movie 3:** polarization-controlled nonlinear metalens with focal distance of 0.5 mm

**Supplementary Method 1:** Sample Fabrication

The samples were fabricated using multilayer e-beam lithography. A 0.15 mm thick borosilicate glass microscope coverslip was used as the substrate. To avoid electron charging on the substrate during SEM, a 3 nm thick chromium layer was initially deposited on the substrate by e-beam evaporation (deposition rate – 0.5 angstroms/s). A 180 nm silica layer was grown by PECVD on top of the Cr film. In the PECVD process, a mixture of gases $SiH_4He(5\%-95\%)/N_2O/N_2$ was used at the respective flows 750/1250/400 sccm. The upper and lower electrodes temperatures were kept at 200 °C and 300 °C respectively, and the total gas pressure at 1200 mTorr, resulting in an average deposition rate of 25 angstroms/s. A 125 nm thick e-beam resist (950k PMMA A2) was spin coated (spin rate 1100 rpm) and pre-baked at 180 °C for 90 s. The patterns were exposed by e-beam lithography using accelerating voltage of 30kV and a beam current of 30pA, at a dose of 400 $\mu C/cm^2$. The resist was then developed by MIBK:IPA (1:3) (30 s) and cleaned with isopropyl alcohol (30 s). In order to fabricate a nearly flat patterned surface of gold antennas, the silica pattern is etched via Inductively Coupled Plasma ($CF_4$ gas, 50 sccm flow, 400W ICP Power and 50 W platen power) for 20 s. A 2 nm thick chromium adhesion layer (evaporation rate – 0.5 angstroms/s) and a 30 nm thick gold layer (evaporation rate 0.7 angstroms/s) were deposited on the patterns by e-beam evaporation. The resist is then lifted off in acetone in an ultrasonic bath. A 180 nm thick silica layer is deposited on the metallic pattern by PECVD and the entire fabrication process is repeated for the additional layers.

**Reference:**

[1]   E. Almeida, G. Shalem, and Y. Prior, "Subwavelength Nonlinear Phase-Control and Anomalous Phase-Matching in Plasmonic Metasurfaces," *arXiv1505.05618,* 2015.